\documentclass[a4paper,12pt]{article}

\title{\LARGE\bf CFTs of SLEs~: the radial case.}
\date{}
\author{}

\newcommand{\tild}{\widetilde}
\newcommand{\vev}[1]{\langle #1 \rangle}
\newcommand{\ket}[1]{| #1 \rangle}
\newcommand{\bra}[1]{\langle #1 |}

\newcommand{\Dtr}{ \mathbb{D}^{\rm tr} }

\newcommand{\monu}{\mathbb{U}}
\newcommand{\monv}{\mathbb{V}}

\def\debut{\begin{eqnarray}}
\def\fin{\end{eqnarray}}

\usepackage[final]{graphicx}
\usepackage{amssymb} 
\usepackage{oldgerm} 

\begin{document}
\maketitle

\vspace{-1.5 truecm}

\centerline{\large Michel Bauer\footnote{Email:
    bauer@spht.saclay.cea.fr} and Denis Bernard\footnote{Member of
    the CNRS; email: dbernard@spht.saclay.cea.fr}} 

\vspace{.3cm}

\centerline{\large Service de Physique Th\'eorique de Saclay}
\centerline{CEA/DSM/SPhT, Unit\'e de recherche associ\'ee au CNRS
\footnote{URA 2306 du CNRS}}
\centerline{CEA-Saclay, 91191 Gif-sur-Yvette, France}


\vspace{1.0 cm}

\begin{abstract}
  We present a relation between conformal field theories (CFT) and
  radial stochastic Schramm-Loewner evolutions (SLE) similar to that
  we previously developed for the chordal SLEs. We construct an
  important local martingale using degenerate representations of the
  Virasoro algebra. We sketch how to compute derivative exponants and
  the restriction martingales in this framework. In its CFT
  formulation, the SLE dual Fokker-Planck operator acts as the
  two-particle Calogero hamiltonian on boundary primary fields and as
  the dilatation operator on bulk primary fields localized at the
  fixed point of the SLE map.
\end{abstract}


\vskip 1.5 truecm


Stochastic Schramm-Loewner evolutions (SLE)
\cite{schramm0,RhodeSchramm,LSW} are random processes adapted to a
probabilistic description of fractal curves or sets growing into
simply connected planar domains $\monu\subset \mathbb{C}$.  They are
expected to provide a rigorous description of two dimensional critical
clusters in their continuous limit.  SLEs depend on a real parameter
$\kappa$.  For $2\leq\kappa<4$, they are conjecturally related to the
$O(n)$ models in their dilute phase with $n=-2\cos(4\pi/\kappa)$, and
to the Fortuin-Kasteleyn clusters of the $Q$-state Potts models with
$Q=4\cos^2(4\pi/\kappa)$ for $4\leq\kappa<8$.  We refer to
refs.\cite{Lawler,cardy} for an introduction to SLEs.

Two classes of SLEs, chordal or radial, have been defined.  The former
-- chordal SLEs -- describe random planar curves joining two points on
the boundary of a domain $\monu$, while the later -- radial SLEs --
describe random curves joining a point on the boundary $\partial
\monu$ to a point in the bulk of $\monu$.  The aim of this note is to
extend to radial SLEs the relation between SLEs and conformal field
theories (CFT), which we developed in the chordal case in
ref.\cite{BaBe}.  Another relation between CFTs and chordal SLEs 
has been presented in ref.\cite{WernerFrie}.

In the first section we shall recall the definition of the radial SLEs
and its basic covariance properties. The CFT formulation of radial
SLEs is given in section 2 in which radial SLEs are viewed as Markov
processes in a completion of the enveloping algebra of a Borel
subalgebra of the Virasoro algebra. This allows us to define a
stochastic evolution operator $\mathcal{A}$, dual to a Fokker-Planck
like operator, which acts on the CFT Hilbert space and whose
geometrical meaning is given. As a consequence, we determine in
section 3 a key local martingale $M_t$ and we show that the dual
Fokker-Planck operator acts as the dilatation operator on spinless
conformal primary fields localized in the bulk ending point of the
radial SLE curves. This agrees with observations made by Cardy in
ref.\cite{cardyN}. The last two sections illustrate the use of the
martingale $M_t$ by re-deriving the derivative exponants computed in
refs.\cite{LSW} and by computing the restriction martingales
\cite{restrictLSW} which code for the response of SLEs to deformations
of the domains in which they are defined.
 
\vskip 1.0 truecm

{\bf 1-Radial SLEs.} The usual description of radial SLEs is by random
curves connecting the point $1$ on the boundary of the unit disc to
the origin. Its study involves a stochastic differential equation,
whose geometric properties play an important role in what follows. If
$\monu$ is any simply connected domain in $\mathbb{C}$, $x_0$ a
boundary point of $\monu$ and $z_*$ an interior point of $\monu$,
there is a unique conformal map from the unit disc to $\monu$ mapping
$1$ to $x_0$ and $0$ to $z_*$. The image of SLE curves by this map
defines a statistical ensemble of random curves in $\monu$, starting
at $x_0$ and ending at $z_*$, which is by
definition radial SLE in $(\monu, x_0,z_*)$. This ensemble is related
to a new stochastic differential equation, which we now describe
geometrically.

Suppose that $f_t(z)$, $t\in[0,T]$, is a familly of
functions solving a stochastic differential equation of the form
$$df_t(z)=dt\, \sigma(f_t(z))+ d\xi_t \, \rho(f_t(z))$$ with
$\xi_t$ a Brownian motion 
with covariance ${\bf E}[\xi_t\,\xi_s]=\kappa\, {\rm min}(t,s)$.
By this we mean that the two functions $\sigma$ and $\rho$ are
holomorphic in some domain $\monu$ and that there is a non empty
domain $\monu_T \subset \monu$ such that $f_t$ maps $\monu_T$ into
$\monu$ and solves the above differential equation for $z \in
\monu_T$. Suppose that $\varphi$ maps $\monu$ conformally to some domain
$\monv$. Then It\^o's formula shows that $f^\varphi_t\equiv \varphi \circ f_t
\circ \varphi^{-1}$ solves the differential equation 
$df^\varphi_t=dt\, \sigma^\varphi
\circ f^\varphi_t + d\xi_t \,\rho^\varphi\circ f^\varphi_t$ 
with $\rho^\varphi \circ \varphi  = \varphi' \rho$ and 
$\sigma^\varphi \circ \varphi =\varphi' \sigma+\frac{\kappa}{2}
\varphi'' \rho^2$. 
These two relations show that 
$$w_{-1} \equiv -\rho(z)\, \partial_z\quad ,\quad 
w_{-2}\equiv \frac{1}{2}\left(-\sigma(z)+
\frac{\kappa}{2}\rho(z)\rho'(z)\right)\partial_z$$ 
transfom as holomorphic vector fields under $\varphi$.

When $\sigma$ and $\rho$ vanish at some point
$z_*\in \monu$, the equation $df_t=dt\, \sigma \circ f_t + d\xi_t \,
\rho \circ f_t$ with $f_0(z)=z$ has a unique solution in some nontrivial
interval $[0,T]$. It satisfies $f_t(z_*)=z_*$ and $f'_t(z_*)\neq 0$.
Inside the space $O_{z_*}$ of germs of holomorphic functions fixing
$z_*$, the subspace $N_{z_*}\equiv \{f \in O_{z_*}, f'(z_*)\neq 0\}$
forms a group for composition, which (anti) acts on $O_{z_*}$ by
$\gamma_f \cdot F \equiv F \circ f$. We may view $f_t$ as a
random process on $N_{z_*}$. Another application of It\^o's formula
shows that $\gamma_{f _t}^{-1} \cdot d \gamma_{f _t} . \cdot F =(dt \,
\sigma + d\xi_t \, \rho)F'+ dt \,\frac{\kappa}{2}\rho^2 F''$, or
equivalently 
\begin{eqnarray}
\gamma_{f _t}^{-1} \cdot d \gamma_{f _t}=dt \,
(-2w_{-2}+\frac{\kappa}{2}w_{-1}^2)- d\xi_t \, w_{-1}.
\label{SLEcov}
\end{eqnarray}
 This equation involves only intrinsic geometric objects. 

To define radial SLE on any domain, 
we only have to choose the vector fields
$w_{-1}$ and $w_{-2}$ appropriately: 
$w_{-1}$ is the generator of conformal motions of $\monu$
fixing $z_*$, and $w_{-2}$ is a  holomorphic vector field, 
unique up to translation by $w_{-1}$, tangent to $\partial \monu$,
fixing $z_*$  but with a pole at $x_0$.

Notice that $g_t \equiv e^{\xi_t w_{-1}}\cdot f_t$ satisfies
$\gamma_{g_t}^{-1} \cdot d \gamma_{g_t}=-2dt \, (e^{-\xi_t
w_{-1}}w_{-2}e^{\xi_t w_{-1}})$, which is an ordinary differential
equation.

We observe that the Lie algebra of $N_{z_*}$ is formally isomorphic to
a completion of a Borel subalgebra of the Virasoro algebra.  In the
sequel, we shall make use of the covariance of radial SLE under
conformal maps to choose $(\monu,x_0,z_*)$ in such a way that $w_{-1}$
and $w_{-2}$ are as simple as possible in terms of Virasoro
generators, so that we can make use of its representation theory and
of conformal field theory efficiently.

In the outer unit disc geometry $\mathbb{D}=\{ z\in\mathbb{C}\, ; |z|\geq 1\}$, 
$w_{-1}=iz\partial_z$ and $2w_{-2}=z \frac{z+1}{z-1}\partial_z$,
and the Loewner equation for the SLE map $g_t=e^{-i\xi_t}f_t$ reads
\cite{schramm0}: 
\begin{eqnarray} 
d\, g_t(z)=- g_t(z)\frac{g_t(z)+U_t}{g_t(z)-U_t}\, dt \quad,\quad
U_t=e^{i\xi_t}.\label{eq:radial}
\end{eqnarray}
with $g_{0}(z)=z$. The SLE hulls $\mathbb{K}_t$ are the sets of points
in $\mathbb{D}$ which have been swallowed: 
$\mathbb{K}_t=\{z\in\mathbb{D}\, ; \tau_z\leq t\}$  
with $\tau_z$ the swallowing time such that
$g_{\tau_z}(z)=U_{\tau_z}$. 
The map $g_t$ is the uniformizing map of
the complement of $\mathbb{K}_t$ in $\mathbb{D}$. 
The SLE curve $\gamma_{[0,\infty)}$, also called the SLE trace, is
reconstructed using $g_t(\gamma(t))=U_t$.

In the upper half plane geometry $\mathbb{H}=\{z\in\mathbb{C}\,;
\Im{\rm m}\, z\geq 0\}$,
$w_{-1}=\frac{1+z^2}{2}\partial_z$ and 
$2w_{-2}=-\frac{1+z^2}{2z}\partial_z$ with $z_*=i$ and $x_0=0$,
so that 
$\tild g_t=(\tild f_t+\eta_t)/(1-\eta_t\,\tild f_t)$ satisfies:
$$
d\, \tild g_t(z) = \frac{1+\tild g_t(z)^2}{2}
\left(\frac{1+\eta_t\, \tild g_t(z)}{\tild g_t(z) - \eta_t} \right)\, dt
\quad,\quad \eta_t=\tan \xi_t/2.
$$
We shall mainly present the results in the case of the disc geometry
but they can easily be translated to the upper half plane geometry.

\vskip 1.0 truecm

{\bf 2- The stochastic evolution operator.}
For making contact with CFT and its operator formalism, 
it is useful to translate the disc by
$-1$ so that the SLE hulls start to be created at $x_0=0$
and grows into $\Dtr$,  the complement in $\mathbb{C}$ of
the unit disc centered at $-1$. 
We define maps $h_t$ by\footnote{The translation by $-1$ is for convenience.
  We could have chosen any other point. The important
  factor is the dilatation by $U^{-1}_t$ which ensures that the tip 
  $\gamma(t)$ of the SLE trace is mapped at any time to the point
  $x_0$ at which it is originally created.} 
$h_t(z)+1=U^{-1}_tg_t(z+1).$
Since we view the hulls as growing outside the unit disc,
both maps $g_t$ and $h_t$ fix the point $z_*=\infty$ at infinity:
$g_t(z_*)=h_t(z_*)=z_*$.
They are normalized there by $g_t(z)=e^{-t}z+O(1)$ and
$h_t(z)=e^{-t-i\xi_t}z + O(1)$.

By the results of \cite{BBpart}, the maps $g_t$ and $h_t$ are associated 
to operators $G_t$ and $H_t$ which implement them in CFT.
Let $L_n$ be the generators of the Virasoro algebra $\mathfrak{vir}$ 
with commutation relations:
$[L_n,L_m]=(n-m)L_{n+m}+\frac{c}{12}n(n^2-1)\delta_{n,-m}$.
The operator $G_t$ belongs to the envoloping algebra of a  Borel
subalgebra of $\mathfrak{vir}$.
From its definition and the radial Loewner equation (\ref{eq:radial}), 
it follows that $G_t$ satisfies:
$$ G^{-1}_t\, d\, G_t = L_0\, dt+ 2 \sum_{n\geq 0} U_t^{n+1}L_{-n-1}\, 
dt.$$
The operator $H_t$ is linked to $G_t$ by
$H_t=e^{-L_{-1}}\, G_t\, e^{i\xi_tL_0}\, e^{L_{-1}}.$
By It\^o calculus,
one finds that $H_t$ satisfies the stochastic equation:
\begin{eqnarray}
H_t^{-1}dH_t = \left( -2W_{-2} +\frac{\kappa}{2}W_{-1}^2\right)\, dt  
- W_{-1}\, d\xi_t 
\label{laradial}
\end{eqnarray} 
with $W_{-1}=i(L_0+L_{-1})$ and  
$W_{-2}=-\frac{1}{2}(L_0+3L_{-1}+2L_ {-2})$. 
Compare with eq.(\ref{SLEcov}).
The stochastic evolution operator ${\cal A}$ is by definition
the drift term in the stochastic equation, eq.(\ref{laradial}): 
\begin{eqnarray}
{\cal A}\equiv -2W_{-2} +\frac{\kappa}{2}W_{-1}^2.
\label{Auniv}
\end{eqnarray}
It may be expressed in terms of the Virasoro generators $L_n$, but those
are associated to the vector fields $\ell_n=-z^{n+1}\partial_z$ which
are not adapted to the geometry of the disc. 
Thus, we change basis and consider the generators $V_n$ associated 
to the vector fields
$v_n=-\frac{i^n}{2}\frac{z^{n+1}}{(z+2)^{n-1}}\partial_z$,
which are the push forward of the $\ell_n$'s by the uniformazing map of
$\Dtr$ onto the upper half plane $\mathbb{H}$. The $V_n$ satisfy the
Virasoro algebra. The first few are: $V_1=\frac{i}{2}L_1$,
$V_0=\frac{1}{2}(L_1+2L_0)$, $V_{-1}=-\frac{i}{2}(L_1+4L_0+4L_{-1})$
and $V_{-2}=-\frac{1}{2}(L_1+6L_0+12L_{-1}+8L_{-2})$. As a
consequence, $W_{-1}=-\frac{1}{2}(V_1+V_{-1})$ and 
$W_{-2}=\frac{1}{4}(V_0+V_{-2})$,
and
\begin{eqnarray}
{\cal A}&=& -\frac{1}{2}(V_0+V_{-2}) + \frac{\kappa}{8}(V_1+V_{-1})^2 
\label{Adisc}\\
&=& \frac{1}{4}(-2V_{-2}+\frac{\kappa}{2}V_{-1}^2)
+(\frac{\kappa-2}{4})V_0 + \frac{\kappa}{8}(V_1^2+2V_{-1}V_1)
\nonumber
\end{eqnarray}

In the upper half plane geometry, the appropriate basis of $\mathfrak{vir}$
are the $L_n$'s and the stochastic evolution operator is
$$ \tild {\cal A}= -\frac{1}{2}(L_0+L_{-2}) +\frac{\kappa}{8}(L_1+L_{-1})^2$$
The basis  $V_n$'s or $L_n$'s correspond to two different
real forms of  $\mathfrak{vir}$.

We need to recall a few basic facts concerning the Virasoro algebra
and its highest weight representations.
For the disc geometry we shall consider highest weight vector
representations defined with respect to the polarization of the
Virasoro algebra associated to the basis $V_n$. So highest weight
vectors $\ket{v}$ are such that $V_n\ket{v}=0$ for $n>0$ and
$V_0\ket{v}=h\ket{v}$. 
For the upper half plane geometry, we shall consider highest weight
representations with respect to the basis $L_n$.

We parametrize the conformal weights by
$$h_{r;s}={[(r\kappa-4s)^2-(\kappa-4)^2]}/{16\kappa}$$ for
$c=1-6{(\kappa-4)^2}/{4\kappa}$. This may also be written in a coulomb
gas representation \cite{nienhuis,FF}, and we shall need it.
 We denote by $2\alpha_0$ the
background charge so that the central charge is
$c=1-12\alpha_0^2$ and the conformal weight of states of coulomb charge
$\alpha$ is $h(\alpha)=\frac{1}{2}\alpha(\alpha-2\alpha_0)$.
The weight $h_{r,s}$ corresponds to the charge $\alpha_{r,s}=\alpha_0
-\frac{r}{2}\alpha_+ -\frac{s}{2}\alpha_-$ with $\alpha_\pm$ the two
screening charges. The correspondance is $\alpha_-=-2\sqrt{2/\kappa}$,
$\alpha_+=\sqrt{\kappa/2}$ and $2\alpha_0=\alpha_++\alpha_-$.

\vskip 1.0 truecm

{\bf 3- Martingales, dilatations and eigenvectors.}
As in the chordal case \cite{BaBe}, a key point is the construction of
an important martingale. It is obtained using degenerate
representations of the Virasoro algebra with null vectors at level two.
We have:
\medskip

{ \it 
Let $\ket{\omega}$ be the highest weight vector in the irreducible
highest weight representation of $\mathfrak{vir}$ of central
charge $c={(6-\kappa)(3\kappa-8)}/{2\kappa}$ and
conformal weight $ h_{1;2}={(6-\kappa)}/{2\kappa}$.
Let $2h_{0;1/2}={(6-\kappa)(\kappa-2)}/{8\kappa}$. Then, 
\begin{eqnarray}
M_t\equiv e^{-2t\, h_{0;1/2}}\, H_t\ket{\omega}
\label{Marti}
\end{eqnarray}
 is a local martingale.}
\medskip

In  particular, by projecting this local martingale on vectors $\bra{v}$ and 
assuming appropriate boundedness conditions, we get that the expectations
$${\bf E}[\,e^{-2t\, h_{0;1/2}}\, \bra{v} H_t\ket{\omega}]$$ are time
independent. 

This result follows from the null vector equation
$(2V_{-2}-\frac{\kappa}{2}V_{-1}^2)\ket{\omega}=0$ which selects the
representation with conformal weight $h_{1;2}$. As a consequence, 
$\mathcal{A}\ket{\omega}=  2h_{0;1/2}\ket{\omega}$ with
$2h_{0;1/2}=(\frac{\kappa-2}{4})\,h_{1;2}$ so that 
$dH_t\ket{\omega}= 2h_{0;1/2}H_t\ket{\omega}dt+
H_tW_{-1}\ket{\omega}d\xi_t$. 

This result may alternatively be formulated in term of the boundary
field $\Psi_{1;2}(x_0)$ creating the state $\ket{\omega}$ at the
origin $x_0$, the point at which the SLE trace starts. 
We have: 
\begin{eqnarray}
\mathcal{A}\cdot \Psi_{1;2}(x_0)= 2h_{0;1/2}\, \Psi_{1;2}(x_0).
\label{Apsi12}
\end{eqnarray}

Other properties emerge when testing the evolution operator against
conformal primary fields. This amounts to 
consider correlation functions with insertions of bulk or
boundary primary fields $\langle\Phi_{\Delta;\bar\Delta}(z,\bar
z)\cdots \Psi_h(x)\cdots \mathcal{A} \ket{\omega}$.
By commuting or deforming contours
 using standard rules of CFT \cite{BPZ,DifBook}, 
the action of $\mathcal{A}=-2W_{-2}+\frac{\kappa}{2}W_{-1}^2$ 
on $\ket{\omega}$ may be traded for an action of 
$$\mathcal{A}^T\equiv
+2W_{-2}+\frac{\kappa}{2}W_{-1}^2$$
 on the bulk and boundary fields. 

Recall that for $V$ a Virasoro generator associated
to a vector field $v(z)$, bulk primary fields $\Phi_{\Delta;\bar
  \Delta}(z,\bar z)$ of dimensions $(\Delta,\bar \Delta)$ 
and boundary primary fields $\Psi_h(x)$ of dimension $h$ satisfy:
\begin{eqnarray}
~[ V ,\Phi_{\Delta,\bar \Delta}(z,\bar z)]&=&
\left( v(z)\partial_z +\Delta v'(z) +  \bar v(\bar z)\partial_{\bar z} +
\bar \Delta \bar v'(\bar z) \right)\Phi_{\Delta,\bar \Delta}(z,\bar z)
\nonumber \\ 
~[ V ,\Psi_h(x)]&=&\left( v(x)\partial_x + 
h\, \Re{\rm e}\,v'(x)\right) \Psi_h(x) \nonumber
\end{eqnarray}
Under global conformal maps $h_t(z)$ implemented in CFT by operators $H_t$, 
primary fields transform as:
\begin{eqnarray*}
H^{-1}_t\, \Phi_{\Delta,\bar\Delta}(z,\bar z)\, H_t &=& h'_t(z)^\Delta\,
 \bar h'_t(\bar z)^{\bar\Delta}\, 
\Phi_{\Delta,\bar\Delta}(h_t(z),\bar h_t(\bar z)) \\
H^{-1}_t\, \Psi_h(x)\, H_t &=& |h'_t(x)|^h \, \Psi_h(h_t(x))
\end{eqnarray*}
As a consequence\footnote{Note that one has to be careful with the
  real form involved in the 
definition of the generator $V$.}, 
the action of $\mathcal{A}^T$ on primary fields is
easy to compute using the explicit expressions of the vector
fields associated to $W_{-2}$ and $W_{-1}$.

For bulk primary fields localized at the point $z_*$ 
fixed by the SLE map, we get the particularly nice result:
\begin{eqnarray}
\mathcal{A}^T\cdot \Phi_{\Delta;\bar\Delta}(z_*,\bar z_*) =
\left( d -\frac{\kappa}{2}\, s^2 \right)\, \Phi_{\Delta;\bar\Delta}(z_*,\bar z_*)
\label{Abulk}
\end{eqnarray}
with $d=\Delta+\bar\Delta$ the scaling dimension and
$s=\Delta-\bar\Delta$ the spin. In other words, $\mathcal{A}^T$, which
may be thought of as a dual of a Fokker-Planck operator, 
acts diagonaly on primary operators
localized at the fixed point. For spinless operators, this action is
simply the dilatation. This is in agreement with the points raised 
in ref.\cite{cardyN}.

On boundary conformal fields, $\mathcal{A}^T$ acts as a second order
differential operator $\mathcal{H}_h$ closely related to the Calogero
hamiltonian.
In the case of the translated disc geometry, with $x=e^{i\theta}-1$
parametrizing the boundary of $\Dtr$, this action reads:
\begin{eqnarray}
\mathcal{A}^T\cdot \Psi_h(x)=
\mathcal{H}_h\cdot \Psi_h(x)\equiv  
\left( \frac{\kappa}{2}\partial_\theta^2 +
  {\rm cotan}\frac{\theta}{2}\partial_\theta 
-\frac{h}{2\sin^2\theta/2}\right) \Psi_h(x)
\label{Abdry}
\end{eqnarray}

The three properties (\ref{Apsi12},\ref{Abulk},\ref{Abdry}) have a
simple consequence: appropriate CFT correlation functions are
eigenfunctions of $\mathcal{H}_h$. For instance,
\begin{eqnarray}
\left(\, \mathcal{H}_h - \epsilon_{\Delta,\bar\Delta}\,\right)\cdot 
\vev{\Phi_{\Delta,\bar\Delta}(z_*,\bar z_*)
\Psi_h(x)\Psi_{1;2}(x_0)} = 0
\label{eigen}
\end{eqnarray}
with $\mathcal{H}_h$ defined above, eq.(\ref{Abdry}), and eigenvalue
\begin{eqnarray}
\epsilon_{\Delta,\bar\Delta}= 2h_{0;1/2} - d +\frac{\kappa}{2}s^2
\label{Edelta}
\end{eqnarray}
with $d=\Delta+\bar\Delta$ and $s=\Delta-\bar\Delta$. 

The simplest case is for $\Psi_h$ the identity operator.
The non vanishing of $\vev{\Phi_{\Delta,\bar\Delta}(z_*,\bar z_*)
\Psi_{1;2}(x_0)}$ then requires $\epsilon_{\Delta,\bar\Delta}=0$,
or equivalently $d=2h_{0;1/2}+\frac{\kappa}{2}s^2$,
which is indeed the fusion rule relation.
This case also provides a simple check of the martingale
$M_t=e^{-2t\, h_{0;1/2}}\,H_t\ket{\omega}$. Indeed, we may compute
$e^{-2t\,h_{0;1/2}}\,\langle\Phi_{\Delta,\bar\Delta}
(z_*,\bar z_*)H_t\ket{\omega}$ by moving $H_t$ to the left so that
$\Phi_{\Delta,\bar\Delta}$ gets transformed by $h_t(z)$. 
Using the normalisation of the SLE map at the fixed point $z_*$ and 
the fusion rule $d=2h_{0;1/2}+\frac{\kappa}{2}s^2$, we get
$exp[{is\xi_t+\frac{\kappa}{2}s^2t}]$ which is a well known martingale 
for the Brownian motion.

Some of the previous results
have a simple interpretation in terms of the $O(n)$ models in the 
dilute phase. Using Coulomb gas techniques, the $L$-leg bulk operators
have been identified \cite{nienhuis,SaleurDuplan} with the bulk operator
$\Phi_{0;L/2}$, and the $L$-leg boundary operators
with $\Psi_{1;L+1}$. Hence the boundary operator
$\Psi_{1;2}$ singled out by the SLE martingale $M_t$ corresponds to a
$1$-leg operator creating a single curve, while the bulk operator
$\Phi_{0;1/2}$ to which it couples corresponds to the termination of a
single curve in the bulk as it should be.

\vskip 1.0 truecm

{\bf 4- Derivative exponants.}  Assume $\kappa>4$.  Derivative
exponants code for the asymptotic behavior of expectations
$f_h(x,t)\equiv {\bf E}[ |h_t'(x)|^h {\bf 1}_{\{\tau_x>t\}}]$, $h\geq
0$, at large time for $x$ on the boundary. In particular $f_0(x,t)$ is
the probability that the point $x$ has not been swallowed by the SLE
trace up to time $t$. As shown in \cite{LSW} using probabilistic
arguments, the time evolution of $f_h(x,t)$ is governed by
$\mathcal{H}_h$, eq.(\ref{Abdry}): $\partial_t f_h(x,t)=
\mathcal{H}_h\cdot f_h(x,t)$. So its large time behavior is dictated
by the eigenstate of $\mathcal{H}_h$ of largest eigenvalue.  We shall
identify this eigenvalue using the martingale
$M_t=e^{-2t\,h_{0;1/2}}H_t\ket{\omega}$.

Indeed, consider as above the projection of the martingale $M_t$ on 
the state created by primary fields localized at the fixed point and
on the boundary: 
$$ 
F_h(x,t)\equiv 
\langle\Phi_{\Delta,\Delta}(z_*,\bar z_*)\Psi_h(x)H_t\ket{\omega}
$$
By construction $e^{-2t\, h_{0;1/2}}F_h(x,t)$, 
$h\geq0$, is a local martingale. 
It may be computed by moving $H_t$ to the left, which conformally
transforms the primary fields. Hence, 
$F_h(x,t)= e^{2\Delta t}\, |h_t'(x)|^h\, F_h(h_t(x),0)$ where we used
the known asymptotic behavior of $h_t$ at the fixed point $z_*$.
As a consequence:
$$ 
{\bf E}[\, |h_t'(x)|^h\, F_h(h_t(x),0)\, ]
= e^{\epsilon_{\Delta,\Delta}\, t}\, F_h(x,0)
$$
with $\epsilon_{\Delta,\Delta}=2h_{0;1/2}-2\Delta$.
This is of course related to the eigen-equation (\ref{eigen}).

Now, as a consequence of the null vector relation
$(2V_{-2}-\frac{\kappa}{2}V_{-1}^2)\ket{\omega}=0$,
the function $F_h(x,0)=
\vev{\Phi_{\Delta,\Delta}(z_*,\bar z_*)\Psi_h(x)\Psi_{1;2}(x_0)}$ 
satisfies a second order differential
equation \cite{BPZ} which depends on $\Delta$.
The primary field $\Phi_{\Delta,\Delta}(z_*,\bar z_*)$ is chosen by
demanding that $F_h(x,0)$ satisfies the same boundary condition as
$f_h(x,t)$. Namely \cite{LSW}, it is single valued when 
$x$ moves along the boundary and it vanishes when $x$
approaches $x_0$ from both sides. 
This selects the conformal weight $\Delta(h)$, 
$$
2 \Delta(h) = \frac{h}{2} + 2h_{0;1/2} + \frac{\kappa}{8}\,\delta_+(h)
$$
with $\delta_\pm(h) =\left[{\kappa-4 \pm \sqrt{(\kappa-4)^2
  + 16h\kappa}}\right]/2\kappa$. With this choice,
$F_h(x,0)=[\sin\theta/2]^{\delta_+(h)}$ in the disc geometry.
This fonction has no node, it is thus the fundamental.
As a consequence, $f_h(x,t)$ decreases exponentially
as $e^{-\lambda(h)\, t}$ with an exponant:
$$ \lambda(h) = 2\Delta(h)-2h_{0;1/2} = 
\frac{h}{2} +\frac{1}{16}\left[\kappa-4 + \sqrt{(\kappa-4)^2
  + 16h\kappa}\,\right]
$$
It of course agrees with ref.\cite{LSW} and with the computations of
ref.\cite{Duplan}, section 12.3, based on 2D quantum gravity.

The dimension $\Delta(h)$ has a simple interpretation in the Coulomb
gas representation. Let $\beta_\kappa=\sqrt{2/\kappa}$ be the charge
of $\Psi_{1;2}$ creating the SLE trace and $\beta$, or
$2\alpha_0-\beta$, be the charge of the boundary operator $\Psi_h$
with $h=\frac{1}{2}\beta(\beta-2\alpha_0)\geq 0$.  Then
$\delta_+(h)=\beta_\kappa\beta$ and
$\delta_-(h)=\beta_\kappa(2\alpha_0-\beta)$ with $\beta>\alpha_0>0$,
reflecting the fact that the fusion relations with $\Psi_{1;2}$ are
linear in terms of Coulomb charges.  $\delta_\pm(h)$ are directly
related to the dimensions of the operators produced by fusing $\Psi_h$
with $\Psi_{1;2}$ since the operator product expansion
$\Psi_h(x)\Psi_{1;2}(x_0)$ behaves as $(x-x_0)^{\delta_+(h)}$ or as
$(x-x_0)^{\delta_-(h)}$ for $x\to x_0$. Hence, the vanishing of
$\vev{\Phi_{\Delta,\Delta}(z_*,\bar z_*)\Psi_h(x)\Psi_{1;2}(x_0)}$ as
$x\to x_0$ demands to represent $\Psi_h$ with the charge $\beta$,
$\beta>\alpha_0$ and not with $2\alpha_0-\beta$.  Let $\alpha=\bar
\alpha$ be the charges of the bulk operator $\Phi_{\Delta,\Delta}$.
Demanding that there are no screening charges in the Coulomb gas
representation of the correlation function
$\vev{\Phi_{\Delta,\Delta}(z_*,\bar z_*)\Psi_h(x)\Psi_{1;2}(x_0)}$
ensures that this function has no monodromy, as we required. Since one
has to compensate the background charge $2\alpha_0$, the absence of
screening charges imposes $2\alpha+\beta+\beta_\kappa= 2\alpha_0$ or
$2\alpha+(2\alpha_0-\beta)+\beta_\kappa=2\alpha_0$.  
Demanding now that this correlation
vanishes as $x$ approaches $x_0$ selects the charge $2\alpha=
2\alpha_0-\beta-\beta_\kappa$. The corresponding scaling dimension
$\alpha(\alpha-2\alpha_0)$ is $2\Delta(h)$.

\vskip 1.0 truecm

{\bf 5- The restriction martingales.}
The martingale $M_t$ may be used to construct the restriction
martingales \cite{restrictLSW}
coding for the influence of deformations of domains on radial SLEs. 
For simplicity we present it in the case of the disc geometry. 
The construction is similar to the one we presented in \cite{BBpart},
so we shall only sketch it.
Let $A$ be a hull in $\Dtr$ and $\phi_A$ be one of the uniformizing map
of its complement onto $\Dtr$ fixing $x_0$.
Given $\phi_A$ and $h_t$, we may write in a unique way
a commutative diagram $\phi_{\widehat A_t}\circ h_t = \widehat h_t\circ
\phi_A$ where $\phi_{\widehat A_t}$ (resp. $\widehat h_t$) uniformizes
the complement of $h_t(A)$ (resp. $\phi_A({\mathbb{K}}_t)$) onto
$\Dtr$ with $\phi_{\widehat A_t}$ fixing $x_0$ and $\widehat h_t$ fixing
$z_*=\infty$. Let as above $H_t$ (resp. $\widehat H_t$) be the operators
implementing $h_t$ (resp. $\widehat h_t$) in CFT. 
Similarly, let $G_{A}$ (resp. $\widehat G_{A_t}$) be those 
implementing $\phi_A$ (resp. $\phi_{\widehat A_t}$). 
Then \cite{BBpart}, 
$$
G^{-1}_{A}\, H_t = Z_t(A)\, \widehat H_t\, \widehat G^{-1}_{A_t}
$$
 with
$$Z_t(A) =\exp \frac{c}{6} \int_0^tds\, (S\phi_{\widehat A_s})(x_0).$$
with $(S\phi)$ the Schwarzian derivative of $\phi$.

By construction $e^{-2t\,h_{0;1/2}}\,
G^{-1}_{A}\,H_t\ket{\omega}$ is a 
local martingale. We may project it on the bulk conformal operator
$\Phi_{0;1/2}$ of dimension $2h_{0;1/2}$ located at the fixed point
$z_*=\infty$.  Computing $\langle \Phi_{0;1/2}(z_*,\bar
z_*)G^{-1}_{A}\,H_t\ket{\omega}$ using the 
commutative diagram yields the martingale 
\footnote{Recall that since $z_*=\infty$, the local coordinate 
around $z_*$ is $1/z$.}~:
$$ M_t(A) \equiv e^{-2t\,h_{0;1/2}} |\widehat h_t'(z_*)|^{-2h_{0;1/2}} 
|\phi_{\widehat A_t}'(x_0)|^{h_{1;2}}\, Z_t(A) $$
Alternatively, since $h_t'(z_*)\phi_{\widehat A_t}'(z_*)=
\widehat h_t'(z_*)\phi_A'(z_*)$ and $|h_t'(z_*)|=e^{-t}$, this reads:
$$ M_t(A)\, |\phi_{A}'(z_*)|^{-2h_{0;1/2}} 
=  |\phi_{\widehat A_t}'(x_0)|^{h_{1;2}}\,
|\phi_{\widehat A_t}'(z_*)|^{-2h_{0;1/2}}\, Z_t(A)
$$
As in \cite{restrictLSW}, this formula may
be used to evaluate the probability that the radial SLE hull at
$\kappa=8/3$ does not touch the hull $A$.
This martingale may be further generalized by projecting
$G^{-1}_{A}\,H_t\ket{\omega}$ on bulk operators with spin
$\Phi_{\Delta,\bar \Delta}(z_*,\bar z_*)$ satisfying the fusion rule
$d=2h_{0;1/2}+\frac{\kappa}{2}s^2$.  

\vskip 1.5 truecm

{\bf Acknowledgements:} 
It is a pleasure to thank John Cardy and Paul Wiegmann for discussions.
Work supported in part by EC contract number
HPRN-CT-2002-00325 of the EUCLID research training network.

\vskip 1.5 truecm


\end{document}